\documentclass{svjour3}                     % onecolumn (standard format)
\smartqed  % flush right qed marks, e.g. at end of proof
\usepackage{graphicx}
\usepackage{amssymb}
\usepackage[mathscr]{eucal}
\usepackage{graphicx}

\usepackage{layout}
\usepackage{lipsum}
\usepackage{enumerate}
\usepackage{amssymb}

\usepackage{mathptmx} 
\usepackage{latexsym}
\usepackage{booktabs}
\usepackage{multirow}
\usepackage{tabularx}
\usepackage{caption}
\usepackage[figuresright]{rotating}
\usepackage{latexsym}
\usepackage{graphicx}
\usepackage[mathscr]{eucal}
\usepackage{amsmath}
\usepackage{mathtools}
\usepackage{booktabs}

\usepackage{array,booktabs,tabularx}

\usepackage{amsmath}
\usepackage{mathtools}

\usepackage[rightcaption]{sidecap}

\usepackage{graphicx} %package to manage images
\graphicspath{ {images/} }

\usepackage{mathptmx}      % use Times fonts if available on your TeX system
\usepackage{latexsym}
\usepackage{booktabs}
\usepackage{multirow}
\usepackage{tabularx}
\usepackage{caption}
\usepackage{latexsym}

 \usepackage{mathptmx}      % use Times fonts if available on your TeX system
\usepackage{latexsym}
\usepackage{booktabs}
\usepackage{multirow}
\usepackage{tabularx}
\usepackage{caption}
\begin{document}

\title{Solution of the  Bethe Equation Through
	the Laplace-Adomian Decomposition Method }

%\titlerunning{Nonlinear oscillations of a point charge}        % if too long for running head

\author{O. Gonz\'alez-Gaxiola, A. Le\'on-Ram\'irez, G. Chac\'on-Acosta }


\institute{O. Gonz\'alez-Gaxiola \at
	Departamento de Matem\'aticas Aplicadas y Sistemas, Universidad Aut\'onoma Metropolitana-Cuajimalpa. Vasco de Quiroga 4871, Santa Fe, Cuajimalpa, 05300, Mexico D.F., Mexico\\
	\email{ogonzalez@correo.cua.uam.mx}  }


\maketitle

\begin{abstract}
The  Bethe equation is a nonlinear differential equation that plays an important role in nuclear physics and a variety of applications related to it, such as the description of the behavior of an energetic particle when it penetrates into matter. Despite its importance, its unusual to find the exact solution to this nonlinear equation in literature and practically all of them are of experimental nature. In this paper, we solve this equation and present a new approach to obtain the solution through the combined use of the  Adomian Decomposition Method and the Laplace Transform (LADM). In addition, we illustrate our approach  solving three examples, in which initial conditions are considered within the typical numerical ranges derived from the applications.  Our results indicate that LADM is  highly accurate  and can be considered a very useful and  valuable method.	

\end{abstract}

\section{Introduction}
\label{Intro}
\noindent Many of the phenomena that arise in the real world can be described by means of nonlinear  partial and ordinary differential equations  and, in some cases,  by integro-differential equations. However, most of the mathematical methods developed thus far are only capable of solving linear differential equations. In the 1980's, George Adomian (1923-1996) introduced a powerful method to solve nonlinear differential equations, known as the Adomian decomposition method (ADM) \cite{ADM-0,ADM-1}. The technique is based on the decomposition of a solution of nonlinear differential equations into a series of functions. Each term of the series is obtained from a polynomial generated by a power series expansion of an analytic function. The Adomian method is very simple in an abstract formulation; however, calculating the polynomials is difficult, which becomes a non-trivial task. This method has been widely used to solve equations that come from nonlinear models, as well as to solve fractional differential equations \cite{Das-1,Das-2,Das-3,Sar}. The  chaotic nature and nonlinearity of other systems, proposed in the past, have been studied through ADM in \cite{Gho}. The advantage of this method is that it solves the problem directly without the need of linearization, perturbation, or any other transformation, and also, requires relatively lesser computational effort compared to most other methods.\\
Quantitative and accurate  information on the penetration of high energy particles through matter, in particular the systematics of energy loss, is a topic of interest in basic science \cite{Mom,Pow,Stern}, medicine and technology \cite{And,Emf,Nass,New,Selt,Stro,Zal}. Until the middle of the past century, studies of charged-particle penetration were stimulated almost exclusively by the needs of fundamental physics research, but applications in other areas gradually became important. The first studies of charged particle penetration were stimulated by experiments on gas discharges toward the end of the 19'th century, but experimental possibilities were greatly enhanced after the discovery of radioactivity, in particular the pioneering work by E. Rutherford and coworkers in the beginning
of the 20'th century. Pioneering theoretical studies by J. J. Thomson and N. Bohr date back to the same time. Subsequently, after the development of quantum mechanics, quantum theory of particle stopping was developed by H. Bethe, F. Bloch, W. H. Barkas, H. H. Andersen and others \cite{Lei,Zie}.\\
\noindent Under certain assumptions, the stopping power in a medium is
given by the relativistic Bethe equation \cite{Bethe-1}. Despite its importance in several physics models, the exact solution of this nonlinear equation have not been obtained.
In the present work, we will use the Adomian  decomposition method in combination with the Laplace transform (LADM) \cite{Waz-Lap} to determine the approximate solution to the Bethe equation. 
We decompose the nonlinear terms of this equation using the Adomian polynomials and then, in combination with the use of the Laplace transform, we obtain an algorithm to solve the problem subject to initial conditions.  Finally, we illustrate our procedure and the quality of the algorithm obtained by solving several numerical examples in which the nonlinear differential equation is solved for different initial conditions.\\
\noindent Our work is divided into several sections. In the ``Adomian Decomposition Method Combined With Laplace Transform'' section, we present, in a brief and self-contained manner, the LADM. Several references are given to delve deeper into the subject and to study its mathematical foundation, which is beyond the scope of the present work. In the ``The Bethe Equation'' section, we present a brief introduction to the model described by the Bethe equation.  In the ``Solution of the Bethe Equation Through LADM'' section, we establish that LADM can be used to solve this equation in a very simple way. In ``Numerical Examples'' section, we show by means of three examples, the quality and precision of our method, comparing the obtained results with existing approximate solutions available in the literature and obtained by other methods. Finally, in the ``Summary and Conclusions"  section, we present the conclusions and implications of this study.

\section{The Adomian Decomposition Method Combined with Laplace Transform}
\label{H-O-0}
\noindent  The ADM is a method  to solve ordinary and nonlinear  differential equations. Using this method, it is possible to express analytic solutions in terms of a series \cite{ADM-1}.
In other words, the method identifies and separates the linear and nonlinear parts of a differential equation. By inverting and applying the highest order differential operator that is contained in the linear part of the equation, it is possible to express the solution in terms of the rest of the equation affected by the inverse operator.  At this point, the solution is proposed through a series
of terms that will be  determined and that will result in the Adomian Polynomials \cite{Waz-0}. The nonlinear part can also be expressed in terms of these polynomials.
The initial (or the border conditions) and the terms that contain the independent variables will be considered as the initial approximation. In this manner, and through recurrence relations, it is possible to find the terms of the series that give the approximate solution of the differential equation. In the next paragraph, we will see how to use the ADM in combination with the Laplace transform (LADM).\\
\noindent \noindent Let us consider the following homogeneous differential equation of first order:
\begin{equation}
\frac{du}{dx}+N(u)=0\label{eq:y1}
\end{equation}
with the initial condition
\begin{equation}
u(0)=u_{0}\label{eq:y2}
\end{equation}
where $u_0$ is a real constant and $N$ is a nonlinear operator acting on the dependent variable $u$ and some of its derivatives.\\  
In general, if we consider the first-order differential operator $L_{x}=\frac{d}{dx}$, then the equation (\ref{eq:y1}) can be  written as 
\begin{equation}
L_{x}u(x)+N(u(x))=0.\label{eq:y3}
\end{equation}
Solving for $L_{x}u(x)$, we have 
\begin{equation}
L_{x}u(x)=-N(u(x))\label{eq:y4}.
\end{equation}
The LADM consists of applying Laplace transform (denoted throughout this paper by $\mathcal{L}$) first on both sides of Eq. (\ref{eq:y4}), thereby obtaining
\begin{equation}
\mathcal{L}\{L_{x}u(x)\}= -\mathcal{L}\{N(u(x))\}.\label{eq:y5}
\end{equation}
An equivalent expression to  (\ref{eq:y5}) is
\begin{equation}
su(s)-u(0)= -\mathcal{L}\{N(u(x))\},\label{eq:y6}
\end{equation}
using the initial condition (\ref{eq:y2}), we have
\begin{equation}
u(s)=\frac{u_0}{s}-\frac{1}{s}\mathcal{L}\{N(u(x))\}.\label{eq:y7}
\end{equation}
Now, applying the inverse Laplace transform to equation (\ref{eq:y7})
\begin{equation}
u(x)=u_{0}-\mathcal{L}^{-1}\big[\frac{1}{s}\mathcal{L}\{N(u(x))\}\big]. \label{eq:y8}
\end{equation}
\noindent The ADM proposes a series of solutions $u(x)$, given by,
\begin{equation}
u(x)= \sum_{n=0}^{\infty}u_{n}(x).\label{eq:y7-1}
\end{equation}
The nonlinear term $N(u)$ is given by
\begin{equation}
N(u)= \sum_{n=0}^{\infty}A_{n}(u_{0},u_{1},\ldots, u_{n})\label{eq:y8-1}
\end{equation}
where   $\{A_{n}\}_{n=0}^{\infty}$ is the so-called Adomian polynomials sequence established in \cite{Waz-0} and \cite{Ba} and, in general, gives us term by term:\\
$A_{0}=N(u_0)$\\
$A_{1}=u_{1}N'(u_0)$\\
$A_{2}=u_{2}N'(u_{0})+\frac{1}{2}u_{1}^{2}N''(u_0)$\\
$A_{3}=u_{3}N'(u_{0})+u_{1}u_{2}N''(u_0)+\frac{1}{3!}u_{1}^{3}N^{(3)}(u_{0})$\\
$A_{4}=u_{4}N'(u_{0})+(\frac{1}{2}u_{2}^{2}+u_{1}u_{3})N''(u_0)+\frac{1}{2!}u_{1}^{2}u_{2}N^{(3)}(u_{0})+\frac{1}{4!}u_{1}^{4}N^{(4)}(u_0)$\\
$ \vdots$.\\
\noindent Other polynomials can be generated in a similar manner. Some other approaches to obtaining Adomian's polynomials can be found in \cite{Duan,Duan1}.\\
\noindent Using  (\ref{eq:y7-1}) and (\ref{eq:y8-1}) in equation (\ref{eq:y8}), we obtain,
\begin{equation}
\sum_{n=0}^{\infty}u_{n}(x)= u_{0}-\mathcal{L}^{-1}\Big[\frac{1}{s}\mathcal{L}\{\sum_{n=0}^{\infty}A_{n}(u_{0},u_{1},\ldots, u_{n})\}\Big].\label{eq:y10}
\end{equation}
From equation (\ref{eq:y10}), we deduce the recurrence formula, which is given as follows:
\begin{equation}
\left\{
\begin{array}{ll}
u_{0}(x)=u_{0},\\
u_{n+1}(x)=-\mathcal{L}^{-1}\Big[\frac{1}{s}\mathcal{L}\{A_{n}(u_{0},u_{1},\ldots, u_{n})\}\Big],\;\; n=0,1,2,\ldots
\end{array}
\right.\label{eq:y11}
\end{equation}
Using  (\ref{eq:y11}), we can obtain an approximate solution of (\ref{eq:y1}), (\ref{eq:y2}) using  
\begin{equation}
u(x)\approx \sum_{n=0}^{k}u_{n}(x),\;\; \mbox{where} \;\; \lim_{k\to\infty}\sum_{n=0}^{k}u_{n}(x)=u(x).\label{eq:y12}
\end{equation}
It is evident that, the Adomian decomposition method, combined with the Laplace transform requires less effort in comparison with the traditional  Adomian decomposition method. This method considerably decreases the volume of calculations. The decomposition procedure of Adomian is  easily set, without requiring the linearization of the problem. With this approach, the solution comes in the form of a convergent series with easily computed components; in many cases, the convergence of this series is very fast and only a few terms are needed to understand how the solutions behave. Convergence conditions of this series are examined by several authors,  mainly in \cite{Y3,Y4,Y1,Y2}. Additional references related to the use of the ADM, combined with the Laplace transform, can be found in \cite{Waz-Lap,Khu,Y} and references therein.

\section{The Bethe Equation}
\label{H-O-1}

\noindent The Bethe equation for the collision stopping
power for incoming charged particles (such as an electron) when it penetrates into matter, is the nonlinear differential equation \cite{Bethe-1}:
\begin{equation}\label{Osc-1}
\frac{du}{dx}+\frac{\ln(u+1)}{u}=0,\quad u(0)=u_{0}.
\end{equation}
\noindent Here $u$ is a dimensionless measure of the kinetic energy, and $x$ is a dimensionless measure of the distance that the particle has penetrated into the matter \cite{Act}. The value $x=0$ corresponds to the surface of the material, where the particle has the high initial energy $u(0)=u_{0}$.\\
\noindent To the best of our knowledge, no exact solution of the nonlinear equation (\ref{Osc-1}) has yet been published; therefore the research work about equation (\ref{Osc-1})  has been intense. \\
\noindent In \cite{He-1} was recently obtained, using an old Chinese algorithm, an approximation to point $R$ in which the kinetic energy $u$ described in (\ref{Osc-1}) is canceled. For technical considerations \cite{37}, the value $R$ in which the kinetic energy is canceled occurs when $u=1$; the value of $R$ as a function of the initial kinetic energy $u_{0}$ calculated in \cite{He-1} is
\begin{equation} \label{hbm}
R=\frac{u_{0}^2}{2(\ln u_{0}-0.55)}.
\end{equation}
Whereas in \cite{Act}, the value of $x_{0}$ as a function of the initial kinetic energy $u_{0}$ was given by the formula
\begin{equation} \label{hbm-1}
R=\frac{u_{0}^2}{2(\ln u_{0}-0.5)}.
\end{equation}
In the following section, we will develop an algorithm using the method described in  the ``Adomian Decomposition Method Combined with Laplace Transform'' section in order to
solve the nonlinear differential equation (\ref{Osc-1}) without resorting to any truncation or linearization. Then we will use that algorithm to solve three problems with initial values of kinetic energy $u_{0}$, included in the typical ranges \cite{Act}, with which we will illustrate that the method used is efficient and highly accurate.

\section{\bf Solution of the Bethe Equation Through LADM}
\label{NLC-ADM}
\noindent Comparing (\ref{Osc-1}) with equation  (\ref{eq:y4}) we have that $L_{x}$ and $N$ becomes:
\begin{equation}
L_{x}u=\frac{d}{dx}u,\;\; Nu=\frac{\ln(u+1)}{u} .\;\; \label{Oper-1}
\end{equation}
\noindent Now, by using equation (\ref{eq:y11}) through the  LADM  method, we recursively obtain
\begin{equation}
\left\{
\begin{array}{ll}
u_{0}(x)=u_{0},\\
u_{n+1}(x)=-\mathcal{L}^{-1}\Big[\frac{1}{s}\mathcal{L}\{A_{n}(u_{0},u_{1},\ldots, u_{n})\}\Big],\;\; n=0,1,2,\ldots
\end{array}
\right.\label{eq:ADM1}
\end{equation}
In addition, the nonlinear term is decomposed as
\begin{equation}
Nu=\frac{\ln(u+1)}{u}=\sum_{n=0}^{\infty}A_{n}(u_{0},u_{1},\ldots, u_{n})
\label{eq:N-1}
\end{equation} 
where $\{A_{n}\}_{n=0}^{\infty}$ is the so-called Adomian polynomials sequence, the terms are calculated according to \cite{Duan,Duan-1,Duan1}. The first few polynomials are given by\\

\begin{equation} \label{A-0}
\begin{split}
A_{0}(u_0)&=\frac{\ln \left(u_0+1\right)}{u_0},
\end{split}
\end{equation}

\begin{equation} \label{A-1}
\begin{split}
A_{1}(u_0,u_1)&=\frac{u_1}{u_0 \left(u_0+1\right)}-\frac{u_1 \ln \left(u_0+1\right)}{u_0^2},
\end{split}
\end{equation}

\begin{equation} \label{A-2}
\begin{split}
A_{2}(u_0,u_1, u_2)&=-\frac{u_1^2}{u_0^2 \left(u_0+1\right)}-\frac{u_1^2}{2 u_0 \left(u_0+1\right)^2}+\frac{u_2}{u_0 \left(u_0+1\right)}\\&+\frac{u_1^2 \ln \left(u_0+1\right)}{u_0^3}-\frac{u_2 \ln \left(u_0+1\right)}{u_0^2},
\end{split}
\end{equation}

\begin{equation} \label{A-3}
\begin{split}
A_{3}(u_0,\ldots, u_3)&=\frac{u_1^3}{u_0^3 \left(u_0+1\right)}+\frac{u_1^3}{2 u_0^2 \left(u_0+1\right)^2}+\frac{u_1^3}{3 u_0 \left(u_0+1\right)^3}-\frac{2 u_2 u_1}{u_0^2 \left(u_0+1\right)}\\&-\frac{u_2 u_1}{u_0 \left(u_0+1\right)^2}+\frac{u_3}{u_0 \left(u_0+1\right)}-\frac{u_1^3 \ln \left(u_0+1\right)}{u_0^4}+\frac{2 u_2 u_1 \ln \left(u_0+1\right)}{u_0^3}\\&-\frac{u_3 \ln \left(u_0+1\right)}{u_0^2},
\end{split}
\end{equation}

\begin{equation} \label{A-4}
\begin{split}
A_{4}(u_0,\ldots, u_4)&=-\frac{u_1^4}{u_0^4 \left(u_0+1\right)}-\frac{u_1^4}{2 u_0^3 \left(u_0+1\right)^2}-\frac{u_1^4}{3 u_0^2 \left(u_0+1\right)^3}-\frac{u_1^4}{4 u_0 \left(u_0+1\right)^4}\\&+\frac{3 u_2 u_1^2}{u_0^3 \left(u_0+1\right)}+\frac{3 u_2 u_1^2}{2 u_0^2 \left(u_0+1\right)^2}+\frac{u_2 u_1^2}{u_0 \left(u_0+1\right)^3}-\frac{2 u_3 u_1}{u_0^2 \left(u_0+1\right)}\\&-\frac{u_3 u_1}{u_0 \left(u_0+1\right)^2}+\frac{u_4}{u_0 \left(u_0+1\right)}-\frac{u_2^2}{u_0^2 \left(u_0+1\right)}-\frac{u_2^2}{2 u_0 \left(u_0+1\right)^2}\\&+\frac{u_1^4 \ln \left(u_0+1\right)}{u_0^5}-\frac{3 u_2 u_1^2 \ln \left(u_0+1\right)}{u_0^4}+\frac{2 u_3 u_1 \ln \left(u_0+1\right)}{u_0^3}\\&+\frac{u_2^2 \ln \left(u_0+1\right)}{u_0^3}-\frac{u_4 \ln \left(u_0+1\right)}{u_0^2},
\end{split}
\end{equation}

\begin{equation} \label{A-5}
\begin{split}
A_{5}(u_0,\ldots, u_5)&=\frac{u_1^5}{u_0^5 \left(u_0+1\right)}+\frac{u_1^5}{2 u_0^4 \left(u_0+1\right)^2}+\frac{u_1^5}{3 u_0^3 \left(u_0+1\right)^3}+\frac{u_1^5}{4 u_0^2 \left(u_0+1\right)^4}\\&+\frac{u_1^5}{5 u_0 \left(U_0+1\right)^5}-\frac{4 u_2 u_1^3}{u_0^4 \left(u_0+1\right)}-\frac{2 u_2 u_1^3}{u_0^3 \left(u_0+1\right)^2}-\frac{4 u_2 u_1^3}{3 u_0^2 \left(u_0+1\right)^3}\\&-\frac{u_2 u_1^3}{u_0 \left(u_0+1\right)^4}+\frac{3 u_3 u_1^2}{u_0^3 \left(u_0+1\right)}+\frac{3 u_3 u_1^2}{2 u_0^2 \left(u_0+1\right)^2}+\frac{u_3 u_1^2}{u_0 \left(u_0+1\right)^3}\\&+\frac{3 u_2^2 u_1}{u_0^3 \left(u_0+1\right)}+\frac{3 u_2^2 u_1}{2 u_0^2 \left(u_0+1\right)^2}+\frac{u_2^2 u_1}{u_0 \left(u_0+1\right)^3}-\frac{2 u_4 u_1}{u_0^2 \left(u_0+1\right)}\\&-\frac{u_4 u_1}{u_0 \left(u_0+1\right)^2}+\frac{u_5}{u_0 \left(u_0+1\right)}-\frac{2 u_2 u_3}{u_0^2 \left(u_0+1\right)}-\frac{u_2 u_3}{u_0 \left(u_0+1\right)^2}\\&-\frac{u_1^5 \ln \left(u_0+1\right)}{u_0^6}+\frac{4 u_2 u_1^3 \ln \left(u_0+1\right)}{u_0^5}-\frac{3 u_3 u_1^2 \ln \left(u_0+1\right)}{u_0^4}\\&+\frac{2 u_4 u_1 \ln \left(u_0+1\right)}{u_0^3}-\frac{3 u_2^2 u_1 \ln \left(u_0+1\right)}{u_0^4}+\frac{2 u_2 u_3 \ln \left(u_0+1\right)}{u_0^3}\\&-\frac{u_5 \ln \left(u_0+1\right)}{u_0^2},
\end{split}
\end{equation}

\begin{equation} \label{A-6}
\begin{split}
A_{6}(u_0,\ldots,u_6)&=\frac{\ln \left(u_0+1\right) u_1^6}{u_0^7}-\frac{u_1^6}{u_0^6 \left(u_0+1\right)}-\frac{u_1^6}{2 u_0^5 \left(u_0+1\right)u^2}-\frac{u_1^6}{3 u_0^4 \left(u_0+1\right)^3}\\&-\frac{u_1^6}{4 u_0^3 \left(u_0+1\right)^4}-\frac{u_1^6}{5 u_0^2 \left(u_0+1\right)^5}-\frac{u_1^6}{6 u_0 \left(u_0+1\right)^6}+\frac{5 u_2 u_1^4}{u_0^5 \left(u_0+1\right)}\\&+\frac{5 u_2 u_1^4}{2 u_0^4 \left(u_0+1\right)^2}+\frac{5 u_2 u_1^4}{3 u_0^3 \left(u_0+1\right)^3}+\frac{5 u_2 u_1^4}{4 u_0^2 \left(u_0+1\right)^4}+\frac{u_2 u_1^4}{u_0 \left(u_0+1\right)^5}\\&-\frac{5 \ln \left(u_0+1\right) u_2 u_1^4}{u_0^6}+\frac{4 \ln \left(u_0+1\right) u_3 u_1^3}{u_0^5}-\frac{4 u_3 u_1^3}{u_0^4 \left(u_0+1\right)}-\frac{2 u_3 u_1^3}{u_0^3 \left(u_0+1\right)^2}\\&-\frac{4 u_3 u_1^3}{3 u_0^2 \left(u_0+1\right)^3}-\frac{u_3 u_1^3}{u_0 \left(u_0+1\right)^4}+\frac{6 \ln \left(u_0+1\right) u_2^2 u_1^2}{u_0^5}+\frac{3 u_4 u_1^2}{u_0^3 \left(u_0+1\right)}\\&+\frac{3 u_4 u_1^2}{2 u_0^2 \left(u_0+1\right)^2}+\frac{u_4 u_1^2}{u_0 \left(u_0+1\right)^3}-\frac{3 \ln \left(u_0+1\right) u_4 u_1^2}{u_0^4}-\frac{6 u_2^2 u_1^2}{u_0^4 \left(u_0+1\right)}\\&-\frac{3 u_2^2 u_1^2}{u_0^3 \left(u_0+1\right)^2}-\frac{2 u_2^2 u_1^2}{u_0^2 \left(u_0+1\right)^3}-\frac{3 u_2^2 u_1^2}{2 u_0 \left(u_0+1\right)^4}+\frac{6 u_2 u_3 u_1}{u_0^3 \left(u_0+1\right)}\\&+\frac{3 u_2 u_3 u_1}{u_0^2 \left(u_0+1\right)^2}+\frac{2 u_2 u_3 u_1}{u_0 \left(u_0+1\right)^3}+\frac{2 \ln \left(u_0+1\right) u_5 u_1}{u_0^3}-\frac{6 \ln \left(u_0+1\right) u_2 u_3 u_1}{u_0^4}\\&-\frac{2 u_5 u_1}{u_0^2 \left(u_0+1\right)}-\frac{u_5 u_1}{u_0 \left(u_0+1\right)^2}+\frac{u_2^3}{u_0^3 \left(u_0+1\right)}+\frac{u_2^3}{2 u_0^2 \left(u_0+1\right)^2}\\&+\frac{u_2^3}{3 u_0 \left(u_0+1\right)^3}+\frac{\ln \left(u_0+1\right) u_3^2}{u_0^3}+\frac{2 \ln \left(u_0+1\right) u_2 u_4}{u_0^3}+\frac{u_6}{u_0 \left(u_0+1\right)}\\&-\frac{\ln \left(u_0+1\right) u_6}{u_0^2}-\frac{\ln \left(u_0+1\right) u_2^3}{u_0^4}-\frac{u_3^2}{u_0^2 \left(u_0+1\right)}-\frac{2 u_2 u_4}{u_0^2 \left(u_0+1\right)}\\&-\frac{u_2 u_4}{u_0 \left(u_0+1\right)^2}-\frac{u_3^2}{2 u_0 \left(u_0+1\right)^2},
\end{split}
\end{equation}

\begin{equation} \label{A-7}
\begin{split}
A_{7}(u_0,\ldots, u_7)&=-\frac{\ln \left(u_0+1\right) u_1^7}{u_0^8}+\frac{u_1^7}{u_0^7 \left(u_0+1\right)}+\frac{u_1^7}{2 u_0^6 \left(u_0+1\right)^2}+\frac{u_1^7}{3 u_0^5 \left(u_0+1\right)^3}\\&+\frac{u_1^7}{4 u_0^4 \left(u_0+1\right)^4}+\frac{u_1^7}{5 u_0^3 \left(u_0+1\right)^5}+\frac{u_1^7}{6 u_0^2 \left(u_0+1\right)^6}+\frac{u_1^7}{7 u_0 \left(u_0+1\right)^7}\\&+\frac{6 \ln \left(u_0+1\right) u_2 u_1^5}{u_0^7}-\frac{6 u_2 u_1^5}{u_0^6 \left(u_0+1\right)}-\frac{3 u_2 u_1^5}{u_0^5 \left(u_0+1\right)^2}-\frac{2 u_2 u_1^5}{u_0^4 \left(u_0+1\right)^3}\\&-\frac{3 u_2 u_1^5}{2 u_0^3 \left(u_0+1\right)^4}-\frac{6 u_2 u_1^5}{5 u_0^2 \left(u_0+1\right)^5}-\frac{u_2 u_1^5}{u_0 \left(u_0+1\right)^6}+\frac{5 u_3 u_1^4}{u_0^5 \left(u_0+1\right)}\\&+\frac{5 u_3 u_1^4}{2 u_0^4 \left(u_0+1\right)^2}+\frac{5 u_3 u_1^4}{3 u_0^3 \left(u_0+1\right)^3}+\frac{5 u_3 u_1^4}{4 u_0^2 \left(u_0+1\right)^4}+\frac{u_3 u_1^4}{u_0 \left(u_0+1\right)^5}\\&-\frac{5 \ln \left(u_0+1\right) u_3 u_1^4}{u_0^6}+\frac{10 u_2^2 u_1^3}{u_0^5 \left(u_0+1\right)}+\frac{5 u_2^2 u_1^3}{u_0^4 \left(u_0+1\right)^2}+\frac{10 u_2^2 u_1^3}{3 u_0^3 \left(u_0+1\right)^3}\\&+\frac{5 u_2^2 u_1^3}{2 u_0^2 \left(u_0+1\right)^4}+\frac{2 u_2^2 u_1^3}{u_0 \left(u_0+1\right)^5}+\frac{4 \ln \left(u_0+1\right) u_4 u_1^3}{u_0^5}-\frac{10 \ln \left(u_0+1\right) u_2^2 u_1^3}{u_0^6}\\&-\frac{4 u_4 u_1^3}{u_0^4 \left(u_0+1\right)}-\frac{2 u_4 u_1^3}{u_0^3 \left(u_0+1\right)^2}-\frac{4 u_4 u_1^3}{3 u_0^2 \left(u_0+1\right)^3}-\frac{u_4 u_1^3}{u_0 \left(u_0+1\right)^4}\\&+\frac{12 \ln \left(u_0+1\right) u_2 u_3 u_1^2}{u_0^5}+\frac{3 u_5 u_1^2}{u_0^3 \left(u_0+1\right)}+\frac{3 u_5 u_1^2}{2 u_0^2 \left(u_0+1\right)^2}+\frac{u_5 u_1^2}{u_0 \left(u_0+1\right)^3}\\&-\frac{3 \ln \left(u_0+1\right) u_5 u_1^2}{u_0^4}-\frac{12 u_2 u_3 u_1^2}{u_0^4 \left(u_0+1\right)}-\frac{6 u_2 u_3 u_1^2}{u_0^3 \left(u_0+1\right)^2}-\frac{4 u_2 u_3 u_1^2}{u_0^2 \left(u_0+1\right)^3}\\&-\frac{3 u_2 u_3 u_1^2}{u_0 \left(u_0+1\right)^4}+\frac{4 \ln \left(u_0+1\right) u_2^3 u_1}{u_0^5}+\frac{3 u_3^2 u_1}{u_0^3 \left(u_0+1\right)}+\frac{3 u_3^2 u_1}{2 u_0^2 \left(u_0+1\right)^2}\\&+\frac{u_3^2 u_1}{u_0 \left(u_0+1\right)^3}+\frac{6 u_2 u_4 u_1}{u_0^3 \left(u_0+1\right)}+\frac{3 u_2 u_4 u_1}{u_0^2 \left(u_0+1\right)^2}+\frac{2 u_2 u_4 u_1}{u_0 \left(u_0+1\right)^3}\\&+\frac{2 \ln \left(u_0+1\right) u_6 u_1}{u_0^3}-\frac{3 \ln \left(u_0+1\right) u_3^2 u_1}{u_0^4}-\frac{6 \ln \left(u_0+1\right) u_2 u_4 u_1}{u_0^4}\\&-\frac{2 u_6 u_1}{u_0^2 \left(u_0+1\right)}-\frac{4 u_2^3 u_1}{u_0^4 \left(u_0+1\right)}-\frac{u_6 u_1}{u_0 \left(u_0+1\right)^2}-\frac{2 u_2^3 u_1}{u_0^3 \left(u_0+1\right)^2}\\&-\frac{4 u_2^3 u_1}{3 u_0^2 \left(u_0+1\right)^3}-\frac{u_2^3 u_1}{u_0 \left(u_0+1\right)^4}+\frac{3 u_2^2 u_3}{u_0^3 \left(u_0+1\right)}+\frac{3 u_2^2 u_3}{2 u_0^2 \left(u_0+1\right)^2}\\&+\frac{u_2^2 u_3}{u_0 \left(u_0+1\right)^3}+\frac{2 \ln \left(u_0+1\right) u_3 u_4}{u_0^3}+\frac{2 \ln \left(u_0+1\right) u_2 u_5}{u_0^3}+\frac{u_7}{u_0 \left(u_0+1\right)}\\&-\frac{\ln \left(u_0+1\right) u_7}{u_0^2}-\frac{3 \ln \left(u_0+1\right) u_2^2 u_3}{u_0^4}-\frac{2 u_3 u_4}{u_0^2 \left(u_0+1\right)}-\frac{2 u_2 u_5}{u_0^2 \left(u_0+1\right)}\\&-\frac{u_3 u_4}{u_0 \left(u_0+1\right)^2}-\frac{u_2 u_5}{u_0 \left(u_0+1\right)^2}.
\end{split}
\end{equation}
Now, recursively using (\ref{eq:ADM1}) with the Adomian polynomials given by the later sequence  $\{A_{n}\}_{n=0}^{\infty}$, we obtain, for a given initial condition
$u_0$:\\
\begin{equation} \label{s-0}
u_{0}(x)=u_0,
\end{equation}
\begin{equation} \label{s-1}
\begin{split}
u_{1}(x)&=-\mathcal{L}^{-1}\Big[\frac{1}{s}\mathcal{L}\{A_{0}(u_{0})\}\Big]=-\frac{x \log \left(u_0+1\right)}{u_0},
\end{split}
\end{equation}
\begin{equation} \label{s-2}
\begin{split}
u_{2}(x)&=-\mathcal{L}^{-1}\Big[\frac{1}{s}\mathcal{L}\{A_{1}(u_{0},u_1)\}\Big]\\&=-\frac{x^2 \ln \left(u_0+1\right)}{2 u_0^3 \left(u_0+1\right)} \Big[-u_0+u_0 \ln \left(u_0+1\right)+\ln \left(u_0+1\right)\Big],
\end{split}
\end{equation}

\begin{equation} \label{s-3}
\begin{split}
u_{3}(x)&=-\mathcal{L}^{-1}\Big[\frac{1}{s}\mathcal{L}\{A_{2}(u_{0},u_1,u_2)\}\Big]\\ &=-\frac{x^3 \ln \left(u_0+1\right) }{6 u_0^5 (u_0+1)^2}\Big[u_0^2+3 u_0^2 \ln ^2\left(u_0+1\right)+6 u_0 \ln ^2\left(u_0+1\right)+3 \ln ^2\left(u_0+1\right)\\& -5 u_0^2 \ln \left(u_0+1\right)-4 u_0 \ln \left(u_0+1\right)\Big],
\end{split}
\end{equation}

\begin{equation} \label{s-4}
\begin{split}
u_{4}(x)&=-\mathcal{L}^{-1}\Big[\frac{1}{s}\mathcal{L}\{A_{3}(u_{0},u_1,u_2,u_3)\}\Big]\\&=-\frac{x^4 \ln \left(u_0+1\right)}{24 u_0^7 \left(u_0+1\right)^3} \Big[-u_0^3+15 u_0^3 \ln ^3\left(u_0+1\right)+45 u_0^2 \ln ^3\left(u_0+1\right)\\ &+45 u_0 \ln ^3\left(u_0+1\right)+15 \ln ^3\left(u_0+1\right)-34 u_0^3 \ln ^2\left(u_0+1\right)-57 u_0^2 \ln ^2\left(u_0+1\right)\\&-25 u_0 \ln ^2\left(u_0+1\right)+15 u_0^3 \ln \left(u_0+1\right)+11 u_0^2 \ln \left(u_0+1\right)\Big] ,
\end{split}
\end{equation}
\begin{equation} \label{s-5}
\begin{split}
u_{5}(x)&=-\mathcal{L}^{-1}\Big[\frac{1}{s}\mathcal{L}\{A_{4}(u_{0},u_1,\ldots,u_4)\}\Big]\\&=-\frac{x^5 \ln \left(u_0+1\right)}{120 u_0^9 \left(u_0+1\right)^4} \Big[u_0^4+105 u_0^4 \ln ^4\left(u_0+1\right)+420 u_0^3 \ln ^4\left(u_0+1\right)\\&+630 u_0^2 \ln ^4\left(u_0+1\right)+420 u_0 \ln ^4\left(u_0+1\right)+105 \ln ^4\left(u_0+1\right)\\&-298 u_0^4 \ln^3\left(u_0+1\right)-772 u_0^3 \ln ^3\left(u_0+1\right)-690 u_0^2 \ln ^3\left(u_0+1\right)\\&-210 u_0 \ln ^3\left(u_0+1\right)+207 u_0^4 \ln ^2\left(u_0+1\right)+319 u_0^3 \ln ^2\left(u_0+1\right)\\&+130 u_0^2 \ln ^2\left(u_0+1\right)-37 u_0^4 \ln \left(u_0+1\right)-26 u_0^3 \ln \left(u_0+1\right)\Big] ,
\end{split}
\end{equation}

\begin{equation} \label{s-6}
\begin{split}
u_{6}(x)&=-\mathcal{L}^{-1}\Big[\frac{1}{s}\mathcal{L}\{A_{5}(u_{0},u_1,\ldots,u_5)\}\Big]\\&=-\frac{x^6 \ln \left(u_0+1\right)}{720 u_0^{11} \left(u_0+1\right)^5} \Big[-u_0^5+945 u_0^5 \ln ^5\left(u_0+1\right)+4725 u_0^4 \ln ^5\left(u_0+1\right)\\&+9450 u_0^3 \ln ^5\left(u_0+1\right)+9450 u_0^2 \ln ^5\left(u_0+1\right)+4725 u_0 \ln ^5\left(u_0+1\right)\\&+945 \ln ^5\left(u_0+1\right)-3207 u_0^5 \ln ^4\left(u_0+1\right)-11310 u_0^4 \ln ^4\left(u_0+1\right)\\&-15372 u_0^3 \ln ^4\left(u_0+1\right)-9450 u_0^2 \ln ^4\left(u_0+1\right)-2205 u_0 \ln ^4\left(u_0+1\right)\\&+3055 u_0^5 \ln ^3\left(u_0+1\right)+7313 u_0^4 \ln ^3\left(u_0+1\right)+6104 u_0^3 \ln ^3\left(u_0+1\right)\\&+1750 u_0^2 \ln ^3\left(u_0+1\right)-954 u_0^5 \ln ^2\left(u_0+1\right)-1402 u_0^4 \ln ^2\left(u_0+1\right)\\&-546 u_0^3 \ln ^2\left(u_0+1\right)+83 u_0^5 \ln \left(u_0+1\right)+57 u_0^4 \ln \left(u_0+1\right)\Big] ,
\end{split}
\end{equation}

\begin{equation} \label{s-7}
\begin{split}
u_{7}(x)&=-\mathcal{L}^{-1}\Big[\frac{1}{s}\mathcal{L}\{A_{6}(u_{0},u_1,\ldots,u_6)\}\Big]\\&=-\frac{x^7 \ln \left(u_0+1\right)}{5040 u_0^{13} \left(u_0+1\right)^6} \Big[u_0^6+10395 u_0^6 \ln ^6\left(u_0+1\right)+62370 u_0^5 \ln ^6\left(u_0+1\right)\\&+155925 u_0^4 \ln ^6\left(u_0+1\right)+207900 u_0^3 \ln ^6\left(u_0+1\right)+155925 u_0^2 \ln ^6\left(u_0+1\right)\\&+62370 u_0 \ln ^6\left(u_0+1\right)+10395 \ln ^6\left(u_0+1\right)-40947 u_0^6 \ln ^5\left(u_0+1\right)\\&-183312 u_0^5 \ln ^5\left(u_0+1\right)-335706 u_0^4 \ln ^5\left(u_0+1\right)-311976 u_0^3 \ln ^5\left(u_0+1\right)\\&-146475 u_0^2 \ln ^5\left(u_0+1\right)-27720 u_0 \ln ^5\left(u_0+1\right)+49640 u_0^6 \ln ^4\left(u_0+1\right)\\&+162636 u_0^5 \ln ^4\left(u_0+1\right)+207403 u_0^4 \ln ^4\left(u_0+1\right)+120582 u_0^3 \ln ^4\left(u_0+1\right)\\&+26775 u_0^2 \ln ^4\left(u_0+1\right)-22714 u_0^6 \ln ^3\left(u_0+1\right)-51800 u_0^5 \ln ^3\left(u_0+1\right)\\&-41328 u_0^4 \ln ^3\left(u_0+1\right)-11368 u_0^3 \ln ^3\left(u_0+1\right)+3775 u_0^6 \ln ^2\left(u_0+1\right)\\&+5388 u_0^5 \ln ^2\left(u_0+1\right)+2037 u_0^4 \ln ^2\left(u_0+1\right)-177 u_0^6 \ln \left(u_0+1\right)-120 u_0^5 \ln \left(u_0+1\right)\Big] ,
\end{split}
\end{equation}

\begin{equation} \label{s-8}
\begin{split}
u_{8}(x)&=-\mathcal{L}^{-1}\Big[\frac{1}{s}\mathcal{L}\{A_{7}(u_{0},u_1,\ldots,u_7)\}\Big]\\&=-\frac{x^8 \ln \left(u_0+1\right)}{40320 u_0^{15} \left(u_0+1\right)^7} \Big[-u_0^7+135135 u_0^7 \ln ^7\left(u_0+1\right)+945945 u_0^6 \ln ^7\left(u_0+1\right)\\&+2837835 u_0^5 \ln ^7\left(u_0+1\right)+4729725 u_0^4 \ln ^7\left(u_0+1\right)+4729725 u_0^3 \ln ^7\left(u_0+1\right)\\&+2837835 u_0^2 \ln ^7\left(u_0+1\right)+945945 u_0 \ln ^7\left(u_0+1\right)+135135 \log ^7\left(u_0+1\right)\\&-605076 u_0^7 \ln ^6\left(u_0+1\right)-3289587 u_0^6 \ln ^6\left(u_0+1\right)-7593561 u_0^5 \ln ^6\left(u_0+1\right)\\&-9468270 u_0^4 \ln ^6\left(u_0+1\right)-6701310 u_0^3 \ln ^6\left(u_0+1\right)-2546775 u_0^2 \ln ^6\left(u_0+1\right)\\&-405405 u_0 \ln ^6\left(u_0+1\right)+891002 u_0^7 \ln ^5\left(u_0+1\right)+3724256 u_0^6 \ln ^5\left(u_0+1\right)\\&+6426369 u_0^5 \ln ^5\left(u_0+1\right)+5667795 u_0^4 \ln ^5\left(u_0+1\right)+2539845 u_0^3 \ln ^5\left(u_0+1\right)\\&+460845 u_0^2 \ln ^5\left(u_0+1\right)-543482 u_0^7 \ln ^4\left(u_0+1\right)-1697378 u_0^6 \ln ^4\left(u_0+1\right)\\&-2071335 u_0^5 \ln ^4\left(u_0+1\right)-1156750 u_0^4 \ln ^4\left(u_0+1\right)-247555 u_0^3 \ln ^4\left(u_0+1\right)\\&+139931 u_0^7 \ln ^3\left(u_0+1\right)+309057 u_0^6 \ln ^3\left(u_0+1\right)+238971 u_0^5 \ln ^3\left(u_0+1\right)\\&+63805 u_0^4 \ln ^3\left(u_0+1\right)-13626 u_0^7 \ln ^2\left(u_0+1\right)-19083 u_0^6 \ln ^2\left(u_0+1\right)\\&-7071 u_0^5 \ln ^2\left(u_0+1\right)+367 u_0^7 \ln \left(u_0+1\right)+247 u_0^6 \ln \left(u_0+1\right)\Big] ,
\end{split}
\end{equation}
$$\vdots . $$
In view of equations (\ref{s-0})-(\ref{s-8}), and considering the equation (\ref{eq:y12}), the approximate solution of the Bethe equation (\ref{Osc-1}) is  
\begin{equation} \label{solser}
u_{\mbox{\tiny LADM}}(x)=u_{0}(x)+u_{1}(x)+u_{2}(x)+\cdots + u_{8}(x).
\end{equation}
In Eq. (\ref{solser}) the approximate solution to the Bethe  equation (\ref{Osc-1}) depends on the initial condition $u_0$. Numerically $u_0$ is typically $10^4$ to $10^6$ \cite{Ev}. 
\section{\bf Numerical Examples}
\noindent In the current section, using the expressions obtained above for the solution of equation (\ref{Osc-1}), we illustrate, with three examples, the effectiveness of LADM to solve the nonlinear Bethe equation. Numerical examples are computed and compared with  the results available in literature. All numerical computations were done with MATHEMATICA software.\\

\begin{figure}[h!]
	\begin{center}
		\includegraphics[width=115mm, height=65mm, scale=1.2]{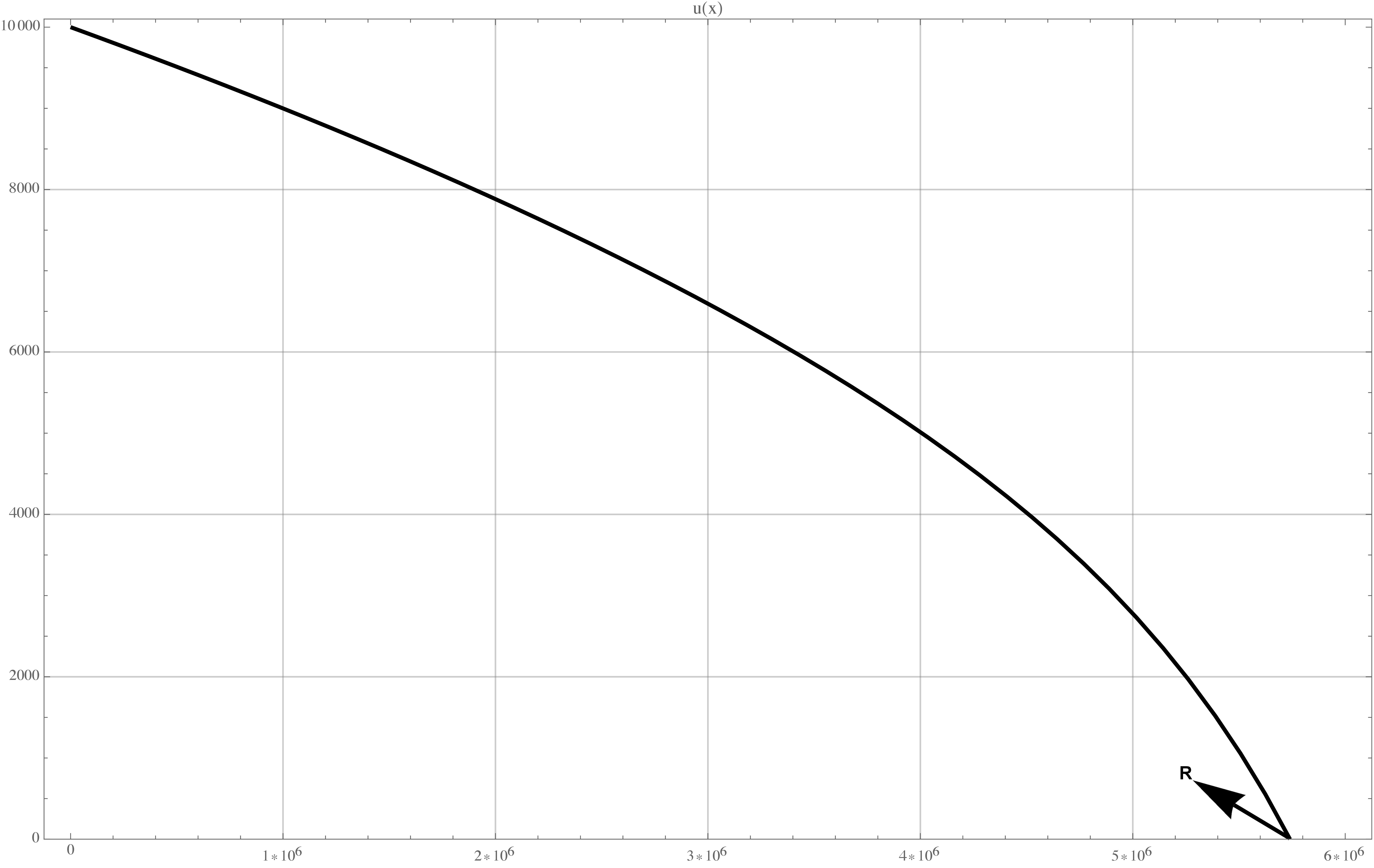}
	\end{center}
	\caption{Plot of the 8-th approximation of $u(x)$ obtained by LADM for $u_0=1\times10^4$ (example 1).\label{fig-1}}
\end{figure}

\noindent {\bf Example 1}\\ For this first example, let us consider the equation of Bethe (\ref{Osc-1}) with the initial condition $u_{0}=1\times 10^{4}$. The approximate solution of (\ref{Osc-1}) is obtained by (\ref{solser}), which in a simplified form is:
\begin{equation} \label{Sol-1}
\begin{split}
u_{\mbox{\tiny LADM}}(x)&=1\times10^{4}-\frac{ \ln (10001)}{1\times10^{4}}\Bigg[x+\frac{x^2 (10001 \ln (10001)-10000)}{20002\times10^{8}}\\&+\frac{x^3 \left(1\times10^{8}+300060003 \ln ^2(10001)-500040000 \ln (10001)\right)}{600120006\times10^{16}}\\&
+\frac{x^4 }{48014401440048\times10^{23}} \Big(-2\times10^{11}+3000900090003 \ln ^3(10001)\\&-6801140050000 \ln ^2(10001)+300022\times10^{7} \ln (10001)\Big)\\& 
+\frac{x^5} {240096014400960024\times10^{32}}\Big(2\times10^{15}+210084012600840021 \ln ^4(10001)\\&-596154413800420000 \ln ^3(10001)+4140638026\times10^{8} \ln ^2(10001)\\&-740052\times10^{11} \ln (10001)\Big)+\frac{x^6}{14407201440144007200144\times10^{40}}\times \\&
\Big(-2\times10^{19}+18909451890189009450189 \ln ^5(10001)\\&-64162623074589004410000 \ln ^4(10001)\\&+61114627220835\times10^{9} \ln ^3(10001)-190828041092\times10^{11} \ln ^2(10001)\\&+1660114\times10^{15} \ln (10001)\Big)\\&
+\frac{x^7}{1008604951220161512060481008\times10^{48}}\Big(2\times10^{23}\\&+2080247711891583118624742079 \ln ^6(10001)\\&-8193066911474398129555440000 \ln ^5(10001)\\&+99312531348301169355\times10^{8} \ln ^4(10001)\\&-45438360826582736\times10^{11} \ln ^3(10001)\\&+755107764074\times10^{15} \ln ^2(10001)-354024\times10^{20} \ln (10001)\Big)\\&
+\frac{x^8}{80696464937222682256934964488064\times10^{56}}\times \Big(-2\times10^{27}\\&
+270459245766160396001758591917027 \ln ^7(10001)\\&-1210810069290157880312936310810000 \ln ^6(10001)\\&+17827489797387160979782169\times10^{8} \ln ^5(10001)\\&-1087303517029013549511\times10^{12} \ln ^4(10001)\\&+27992381617954761\times10^{16}\ln ^3(10001)\\&-2725581674142\times10^{19} \ln ^2(10001)+7340494\times10^{23} \ln (10001) \Big)\Bigg].
\end{split}
\end{equation}
Figure \ref{fig-1} shows the graph of the solution of the Bethe equation obtained through LADM for the initial condition $u_0=1\times10^4$, in addition we can find the physical range $R$ of the particle with initial energy $u_0$, which is obtained when $u_{LADM}\approx 1$ as we can see in \cite{Act}.

\begin{figure}[h!]
	\begin{center}
		\includegraphics[width=115mm, height=65mm, scale=1.0]{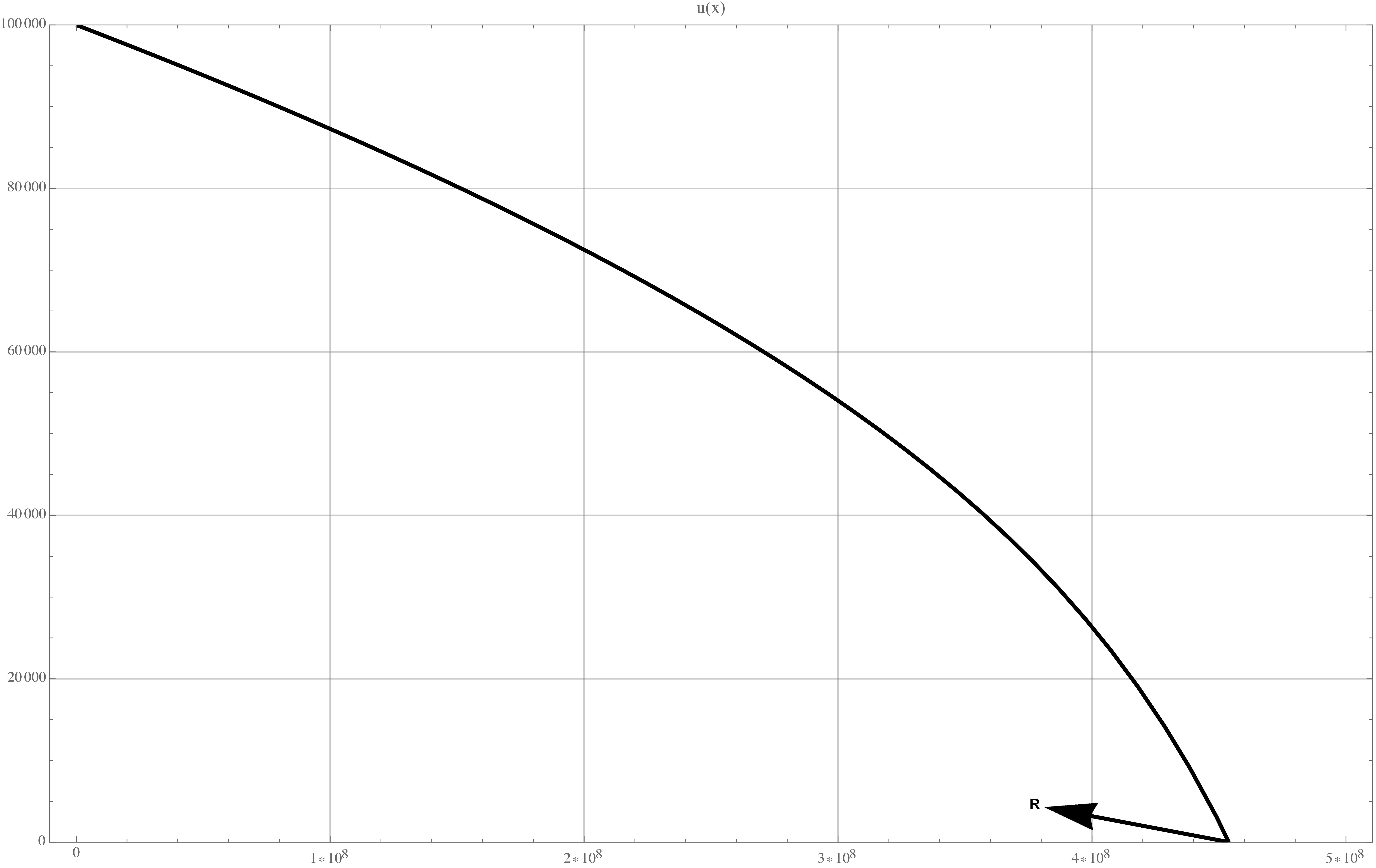}
	\end{center}
	\caption{Plot of the 8-th approximation of $u(x)$ obtained by LADM for $u_0=1\times10^5$ (example 2). \label{fig-2}}
\end{figure}

\noindent {\bf Example 2}\\ 
In this second example, we will consider the Bethe equation (\ref{Osc-1}) with the initial condition $u_{0}=1\times 10^5$. The approximate solution of (\ref{Osc-1}) is obtained by (\ref{solser}), which is given by: 

\begin{equation} \label{Sol-2}
\begin{split}
u_{\mbox{\tiny LADM}}(x)&=1\times10^{5}-\frac{ \ln (100001)}{1\times10^{5}}\Bigg[x+\frac{x^2 (100001 \ln (100001)-100000)}{200002\times10^{10}}\\&+\frac{x^3 \left(1\times10^{10}+30000600003 \ln ^2(100001)-500004\times10^{5} \ln (100001)\right)}{60001200006\times10^{20}}\\&
+\frac{x^4 }{48001440014400048\times10^{29}} \big(-2\times10^{14}+3000090000900003 \ln ^{3}(100001)\\& -68001140005\times10^{5} \ln ^2(100001)+3000022\times10^{9} \ln (100001)\big)\\& 
+\frac{x^5} {2400096001440009600024\times10^{40}}\Big(2\times10^{19}\\&+2100084001260008400021 \ln ^4(100001)-59601544013800042\times10^{5} \ln ^3(100001)\\&+414006380026\times10^{10} \ln ^2(100001)-7400052\times10^{14} \ln (100001)\Big)\\&
+\frac{x^6}{1440072001440014400072000144\times10^{50}}\times 
\Big(-2\times10^{24}\\&+1890094501890018900094500189 \ln ^5(100001)\\&-64142262030744189000441\times10^{5} \ln ^4(100001)\\&+61101462612208035\times10^{11} \ln ^3(100001)\\&-19080280401092\times10^{14} \ln^2(100001)+16600114\times10^{19} \ln(100001)\Big)\\&
+\frac{x^7}{1008060481512020160151200604801008\times10^{60}}\Big(2\times10^{29}\\&+2079124743118541580311851247402079 \ln ^6(100001)\\&-8189766630714182395492950554400000 \ln ^5(100001)\\&+992832527614808411645355\times10^{10} \ln ^4(100001)\\&-45429036008265622736\times10^{14} \ln ^3(100001)\\&+75501077604074\times10^{19} \ln ^2(100001)-3540024\times10^{25} \ln (100001)\Big)\\&
+\frac{x^8}{806456449693468224282241693445644808064\times10^{70}}\times \Big(-2\times10^{34}\\&
+2702889194675764595445950675688918927027 \ln ^7(100001)\\&-12102177932587311366740267093558108100000 \ln ^6(100001)\\&+1782078486405285135640796992169\times10^{10} \ln ^5(100001)\\&-10869979479742693135049511\times10^{15} \ln ^4(100001)\\&+27986818118779432761\times10^{20} \ln ^3(100001)-272523816614142\times10^{24} \ln ^2(100001)\\&+73400494\times10^{29} \ln(100001) \Big)\Bigg].
\end{split}
\end{equation}
Figure \ref{fig-2} shows the graph of the solution of the Bethe equation obtained through LADM for the initial condition $u_0=1\times10^5$, in addition we can find the physical range $R$ of the particle with initial energy $u_0$, which is obtained when $u_{LADM}\approx 1$, see \cite{Act}.

\begin{figure}[h!]
	\begin{center}
		\includegraphics[width=115mm, height=65mm, scale=1.0]{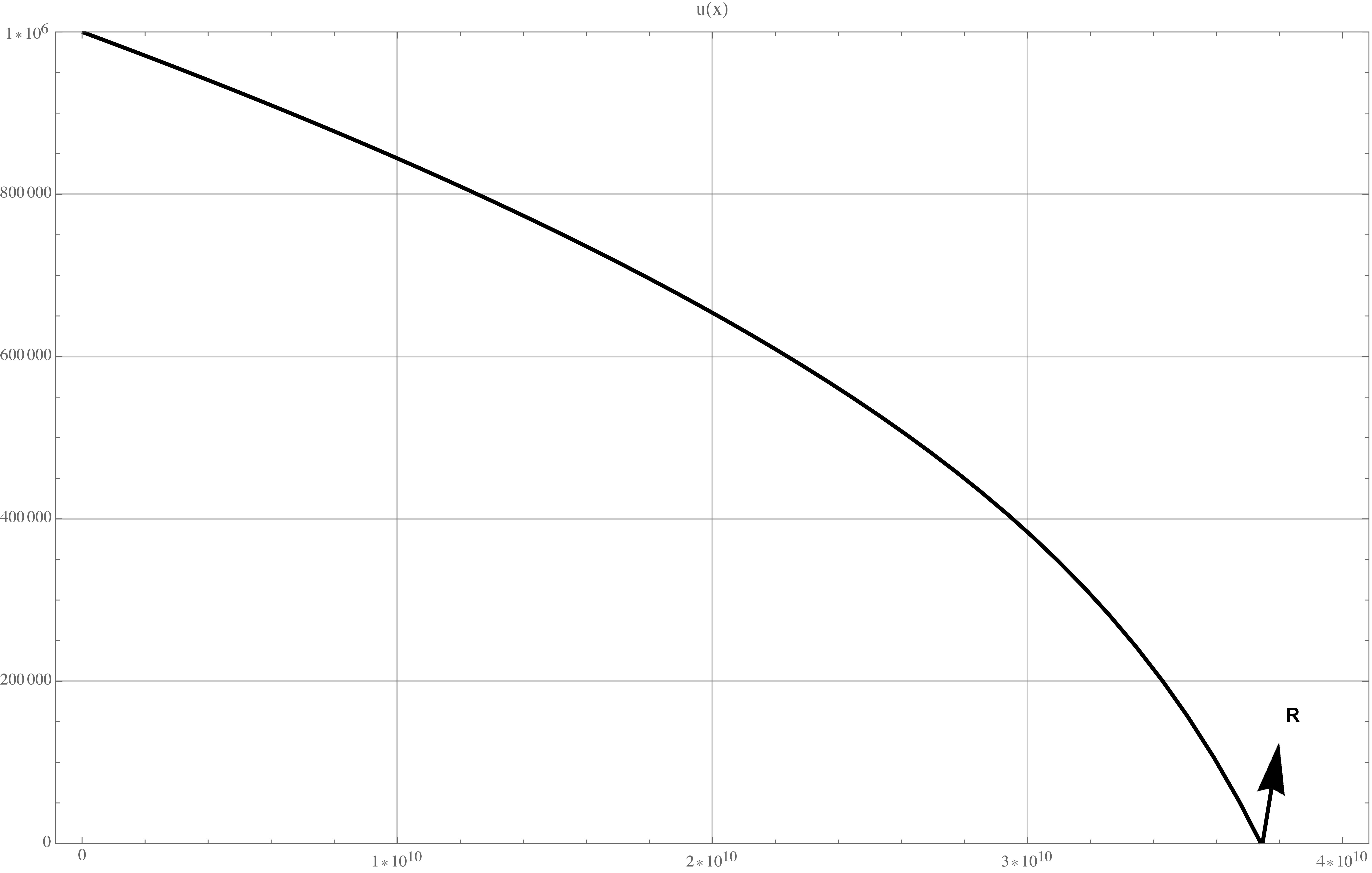}
	\end{center}
	\caption{Plot of the 8-th approximation of $u(x)$ obtained by LADM for $u_0=1\times10^6$ (example 3). \label{fig-3}}
\end{figure}

\noindent {\bf Example 3}\\ 
In this example, we will consider the Bethe equation (\ref{Osc-1}) with the initial condition $u_{0}=1\times 10^6$. The approximate solution of (\ref{Osc-1}) is obtained by (\ref{solser}), which is given by:

\begin{equation} \label{Sol-3}
\begin{split}
u_{\mbox{\tiny LADM}}(x)&=1\times10^{6}-\frac{ \ln (1000001)}{1\times10^{6}}\Bigg[x+\frac{x^2 (1000001 \ln (1000001)-1000000)}{2000002\times10^{12}}\\&+\frac{x^3 \left(1\times10^{12}+3000006000003 \ln ^2(1000001)-5000004\times10^{6} \ln (1000001)\right)}{6000012000006\times10^{24}}\\&
+\frac{x^4 }{48000144000144000048\times10^{35}} \Big(-2\times10^{17}\\&+3000009000009000003 \ln ^3(1000001)-6800011400005\times10^{6} \ln ^2(1000001)\\&+30000022\times10^{11} \ln(1000001)\Big)\\& 
+\frac{x^5} {24000096000144000096000024\times10^{48}}\Big(2\times10^{23}\\&+21000084000126000084000021 \ln ^4(1000001)\\&-59600154400138000042000000 \ln ^3(1000001)\\&+41400063800026\times10^{12} \ln ^2(1000001)-74000052\times10^{17} \ln (1000001)\Big)\\&
+\frac{x^6}{144000720001440001440000720000144\times10^{60}}\times \\&
\Big(-2\times10^{29}+189000945001890001890000945000189 \ln ^5(1000001)\\&-641402262003074401890000441\times10^{6} \ln ^4(1000001)\\&+61100146260122080035\times10^{13} \ln ^3(1000001)\\&-1908002804001092\times10^{17} \ln ^2(1000001)+166000114\times10^{23} \ln (1000001)\Big)\\&
+\frac{x^7}{1008006048015120020160015120006048001008\times10^{72}}\Big(2\times10^{35}\\&+2079012474031185041580031185012474002079 \ln ^6(1000001)\\&-8189436662467141262395229295005544000000 \ln ^5(1000001)\\&+9928032527241480624116405355\times10^{12} \ln ^4(1000001)\\&-45428103600082656022736\times10^{17} \ln ^3(1000001)\\&+7550010776004074\times10^{23} \ln ^2(1000001)-35400024\times10^{30} \ln (1000001)\Big)\\&
+\frac{x^8}{8064056448169344282240282240169344056448008064\times10^{84}}\\&\times \Big(-2\times10^{41}
+27027189189567567945945945945567567189189027027 \ln ^7(1000001)\\&-121015857918918714093655340262509355081081\times10^{6} \ln ^6(1000001)\\&+178201144852485274933559507969092169\times10^{12} \ln ^5(1000001)\\&-108696739476014267231350049511\times10^{18} \ln ^4(1000001)\\&+27986261811447794212761\times10^{24} \ln ^3(1000001)\\&-27252038166014142\times10^{29} \ln ^2(1000001)+734000494\times10^{35} \ln (1000001) \Big)\Bigg].
\end{split}
\end{equation}
Figure \ref{fig-3} shows the graph of the solution of the Bethe equation obtained through LADM for the initial condition $u_0=1\times10^6$, in addition we can find the physical range $R$ of the particle with initial energy $u_0$, which is obtained when $u_{LADM}\approx 1$.

\noindent  The results obtained in the three previous examples are shown in Table  \ref{tab:Z1}, in which the values of the physical range R obtained using the results reported in \cite{He-1} and \cite{Act} are compared for different values of $u_0$. All numerical work was performed using the Mathematica software package.
\begin{table}[h!]
	\tabcolsep=0.15cm
	\begin{tabular}{cccccc}
		%	\toprule
		\multicolumn{6}{c}{} \\
		\cmidrule(r){1-6}
		$u_0$  & $R$ by present method &$R_1$ obtained in \cite{He-1} & $R_{2}$ obtained in \cite{Act} &$\frac{ \vert R_{1}-R_{2}\vert}{R_2}\times 100$&$\frac{ \vert R-R_{2}\vert}{R_2}\times 100$\\
		\midrule
		$1\times10^4$&$5741211.4$&$5773445.1$&$5740303.8$&$0.57\% $&$0.015\%$\\
		$1\times10^5$&$4.5401472\times10^8$&$7.5608264\times10^8$&$4.5401197\times10^8$&$0.27\% $&$0.0006\%$\\
		$1\times10^6$&$3.7550206\times10^{10}$&$3.7691727\times10^{10}$&$3.7550194\times10^{10}$&$0.37\% $&$0.00003\%$\\
		\bottomrule
	\end{tabular}
	\caption{Table of comparison of the physical range R obtained by the present method and those provided by the formulas (\ref{hbm}) and (\ref{hbm-1}).}
	\label{tab:Z1}   
\end{table}

\vspace{0.1in}

\noindent In Table \ref{tab:Z1} we can see that our approximation made in the three previous examples to the solution of the Bethe equation is very good, since by comparing the rank $R$ that is obtained for the same initial kinetic energies $u_0$ in the only known recent works, that have addressed the problem, we can see that our approach is suitable since the maximum error is $0.015\%$. \\
We remark that, as these examples demonstrate, the Adomian  decomposition method in combination with the Laplace transform avoids several difficulties in the calculation including massive computational work are
required, e.g., by discretization techniques, in determining the approximate analytic solution.

\section{Summary and Conclusions}

\noindent Very few exact solutions of the Bethe equation were known in the literature and practically all of them are of experimental nature. In this paper, we have obtained accurate approximations for the Bethe nonlinear differential equation solution using the Adomian decomposition method in combination with the Laplace transform, illustrating, in this way, the use of LADM in the solution of nonlinear  differential equations. We have chosen the  Bethe equation for its importance in physics  as it describes the interaction between radiation and matter.\\
\noindent In order to show the accuracy and efficiency of our method,  we have solved  three examples, comparing our results with the only approximations that are known not from the Bethe equation itself but in the calculation of the physical range reached by a particle whose kinetic energy evolves according to said equation, which can be seen in the Table \ref{tab:Z1}. Our results show that LADM produces highly accurate solutions in complicated nonlinear problems.
\noindent We therefore,  conclude that the Laplace-Adomian decomposition method is a notable  non-sophisticated powerful tool that produces high quality approximate solutions for nonlinear differential equations using simple calculations and that attains  converge with only few terms. All numerical work and graphics were performed with the Mathematica software package.

%\begin{figure} [h!]
%	\begin {center}
%	\includegraphics[width=0.3\textwidth]{Cont-1.jpg}\hspace{0.3in}
%	\includegraphics[width=0.3\textwidth]{Cont-2.jpg}
%	\caption{\small  Illustration of a glioblastoma tumor in the parietal lobe. }
%	\end {center}
%\end{figure}

%\begin{figure}[h!]
%	\begin{center}
%		\includegraphics[width=80mm, height=57mm, scale=1.0]{Re-1.jpg}
%	\end{center}
%	\caption{\tiny{Plot of the real part of the approximate solution $u_{LADM}$ versus the real part of the $u_{exact}$}. \label{fi1}}
%\end{figure}

%\begin{figure}[h!]
%	\begin{center}
%		\includegraphics[width=80mm, height=48mm, scale=0.9]{Im-1.jpg}
%	\end{center}
%	\caption{\tiny{Plot of the imaginary part of the approximate solution $u_{LADM}$ versus the imaginary part of the $u_{exact}$}. \label{fi2}}
%\end{figure}

%\begin{figure}[h!]
%	\begin{center}
%		\includegraphics[width=80mm, height=50mm, scale=0.8]{CompRe.jpg}
%	\end{center}
%	\caption{{\tiny Graph of real part of $u_{LADM}$ versus real part of $u_{ex}$ for $t =1, 2, 3, 4, $ and $ 5 $}.\label{fi3}}
%\end{figure}

%\begin{figure}[h!]
%	\begin{center}
%		\includegraphics[width=80mm, height=50mm, scale=0.8]{CompIm.jpg}
%	\end{center}
%	\caption{{\tiny Graph of imaginary part of $u_{LADM}$ versus imaginary part of $u_{ex}$ for $t =1, 2, 3, 4, $ and $ 5 $}.\label{fi4}}
%\end{figure}

%\section*{Acknowledgments}


\end{document}